# Constructing a Multiple-Choice Assessment For Upper-Division Quantum Physics From An Open-Ended Tool

Homeyra Sadaghiani*, John Miller*, Steven Pollock†, and Daniel Rehn†

*Department of Physics and Astronomy, Cal Poly Pomona, Pomona, CA 91768
† Department of Physics, University of Colorado, Boulder, CO 80309

**Abstract:** As part of an ongoing investigation of students' learning in upper-division quantum mechanics, we needed a high-quality conceptual assessment instrument for comparing outcomes of different curricular approaches. The 14 item open-ended Quantum Mechanics Assessment Tool (QMAT) was previously developed for this purpose. However, open-ended tests require complex scoring rubrics, are difficult to score consistently, and demand substantial investment of faculty time to grade. Here, we present the process of converting open-ended questions to multiple-choice (MC) format. We highlight the construction of effective distractors and the use of student interviews to revise and validate questions and distractors. We examine other elements of the process, including results of a preliminary implementation of the MC assessment given at Cal Poly Pomona and CU Boulder.

## INTRODUCTION

As instructors and researchers, we seek reliable tools to effectively measure student learning. Such tools allow us to study the efficacy of different curricula or classroom activities, and help to identify common student difficulties. By making student thinking apparent, a well-developed instrument can help guide efforts to systematically improve instruction. A number of assessment tools are available to physics instructors for introductory level courses [1-3], which have been instrumental in supporting and evaluating transformed pedagogies. In recent years, a number of assessment tools that focus on intermediate and upper division undergraduate physics [4-5] have also been developed.

In quantum mechanics (QM), a growing body of research into student difficulties has led to the development of a variety of assessment tools. Existing tools focus on particular issues such as measurement [6] and visualizations [7]. The Quantum Mechanics Conceptual Survey (QMCS) offers a broad assessment of QM topics, but is designed for sophomore level modern physics [8]. The Quantum Mechanics Survey (QMS) spans a variety of important topical areas in one spatial dimension [9], and is broadly appropriate for upper-division QM, albeit with some emphasis on formalism.

The Quantum Mechanics Assessment Tool (QMAT) [10] is an open-ended instrument developed at CU Boulder to address faculty-consensus learning goals and incorporate findings on student difficulties in advanced undergraduate quantum mechanics [11-17]. However, the QMAT suffers from a complicated scoring rubric, with correspondingly limited validation studies. There are a variety of known difficulties associated with reliable scoring of open-ended questions [18, 19], and such issues have restricted the usefulness and transferability of the QMAT within and across institutions.

We are currently in the process of constructing a conceptually focused multiple-choice (MC) tool by building upon the original QMAT. MC tests have some advantages over open-ended tests; e.g., they can be easily and accurately graded, and are less ambiguous to validate. High quality MC tests with proper distractors have a long tradition of providing diagnostics of student difficulties, evaluating teaching methods and comparing curricula. The QMAT is a valuable tool capable of leading us to our long-term goal of have an instrument capable of equally assessing outcomes from modern (discrete spin half basis) and traditional approaches (continuous single particle basis) to teaching quantum mechanics.

The construction of a reliable and comprehensive conceptual instrument is a nontrivial, multi-step process. We are guided by elements of classical test theory [20] as a suitable approach for constructing a conceptual multiple-choice test. In this paper, we discuss our ongoing process of constructing and refining MC distractors from existing QMAT questions and present preliminary tests of item validity using expert and student feedback.

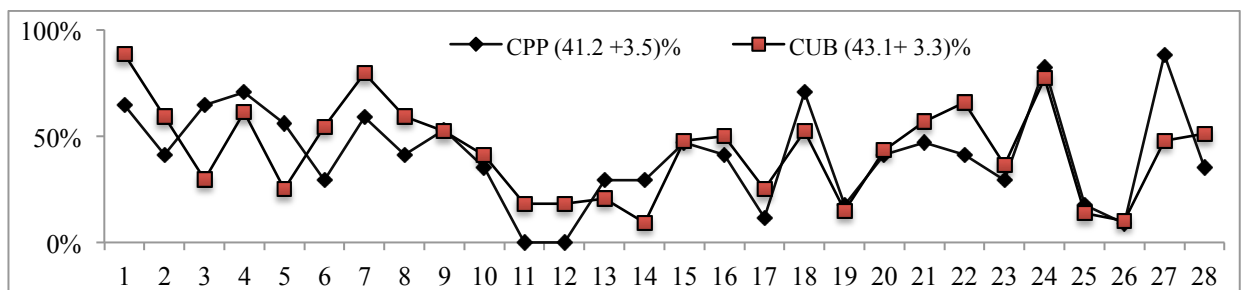

**Figure 1.** Item difficulty distribution for questions from two implementations of the first version of MC test.

## CONSTRUCTING MC ITEMS

An initial set of MC distractors was developed using student responses to the open-ended QMAT. Student responses from Cal Poly Pomona (CPP; N=19) and CU Boulder (CUB; N=53) were used in the construction of distractors. Students' written responses were categorized into groups of similar ideas that formed distractors for the MC version. In addition to using student responses to the open-ended QMAT, we examined existing literature [11-17] to ensure congruence of known student difficulties in the chosen distractors. In some cases, student correct and incorrect responses produced a broad spectrum of plausible distractors and keys. In the first MC version, we allowed students to choose more than one correct response. This approach allowed us to consider a broader array of student-generated ideas as distractors at early stages of the development. However, in the ensuing version, the less popular distractors were taken out and correct responses were integrated into one single choice to allow for standard MC format grading and to decrease the overall difficulty of the test.

The first MC version consisted of 28 questions and was given to a total of 61 students [(N = 17) at CPP and (N = 44) at CUB]. The average scores for individual questions were strikingly similar for CPP and CUB, suggesting that the relative item difficulty of the MC test for students from different populations is similar (**Fig. 1**). The mean scores for the two institutions were comparable but very low, indicating that the test is challenging. We used these data and student interviews to improve the questions and distractors for questions with very small or large item difficulties. To demonstrate the mechanism by which QMAT questions were converted to MC and iterated upon through the use of data, we discuss some rationale and issues for revisions of two representative examples.

### Example 1: Time Oscillation of Position Probability Density

The original open-ended question six (Q6) asked students to invent examples of a quantum state whose position probability density is (a) time independent, (b) time dependent, and then (c) explain how they could change their answer to part (b) to make the position probability density change more rapidly. The overall average correct response to the open-ended question was 37% [(a) 44%, (b) 39%, (C) 28%] at CPP (N=19) and 54% [(a) 66%, (b) 59%, (C) 38%] at CUB (N=53).

**Fig. 2** shows two MC questions we constructed to address some of the learning goals of Q6. The item analysis of these questions in **Table 1** shows that some distractors were more popular than others. Unpopularity of option (a) and (e) in MC-Q16 provided valuable feedback for modifying the distractors on the next iteration of the test. While the item difficulty of MC-Q16 is in an ideal range (~50% correct), MC-Q17 is overly challenging for the students. Thus, we focused on modifying MC-Q17. Below, we discuss our rationale, along with student feedback from interviews.

---

**Q16.** Which of the following statements are always true for a particle in a one-dimensional infinite square well?
 a. The position expectation value is time-dependent for energy eigenstates.
 b. The position expectation value is time-independent for energy eigenstates.
 c. The position expectation value is always time-independent.
 d. The position expectation value is always time-dependent.
 e. More than one of the above statements is always true.

**Q17.** For a particle in a one-dimensional infinite square well, which state will have the fastest variation in time for the position probability density? (The state $\psi_n$ corresponds to an energy of $E_n$)
 a. $\psi_1$
 b. $\psi_4$
 c. $\frac{1}{\sqrt{2}}(\psi_2 + \psi_3)$
 d. $\frac{1}{\sqrt{2}}(\psi_1 - \psi_3)$
 e. All of these states have time-independent position probability densities.

**Figure 2.** The first version of a set of MC questions corresponding to elements of the open-ended QMAT Q6.

|     |   | Q16 (%) | Q17 (%) |
|-----|---|---------|---------|
|     | a | 0       | 6       |
|     | b | **41**  | 25      |
| CPP | c | 24      | 32      |
|     | d | 35      | **12**  |
|     | e | 0       | 19      |
|     | a | 2%      | 0       |
|     | b | **50**  | 20      |
| CUB | c | 34      | 27      |
|     | d | 14      | **25**  |
|     | e | 0       | 23      |

**Table 1**. Distribution of students' responses for MC questions related to open-ended QMAT Q6.

### Interviews on MC-Q17

We conducted thirteen individual interviews (7 at CPP, 6 at CUB) using a think-aloud method [21]. In student interviews on MC-Q17, we observed that all students, even those without a clear understanding of the question, dismissed option (a) stating, e.g.:

"$\psi_1$ is a ground state so has no oscillations."

This was not the case for $\psi_4$ in option (b). In fact, a student who did not initially dismiss $\psi_4$ sketched an oscillation in the air for $\psi_4$ by gesturing his right index finger up and down and just drew a hump for $\psi_1$. Another student did not distinguish between the wave function frequency and time oscillation frequency of the position probability density of a superposition state, and after selecting $\psi_4$ stated:

"... the fastest variation in time will be given by the highest energy level."

Thus, from student interview data, $\psi_1$ was being dismissed for being the lowest energy and its particular shape and not for being a stationary eigenstate.

Some students also made use of their classical wave knowledge in answering this question. For example, a student who dismissed options (a) and (b) had difficulty choosing between options (c) and (d), stating:

"My mind goes back to acoustic waves. When you have two different frequencies, from the difference in the frequencies you hear beats. So, I would say that the greatest variation would be for energies more far apart."

Such an explanation (that led to the correct choice) was not mentioned in the course. Nevertheless, it reveals a range of different analogies and resources students use in answering quantum physics questions. We are investigating the ways in which observations such as this could further guide us in improving the MC distractors. To address the low discrimination of MC-Q17, we will be testing new questions guided by interview data in the next version, which now looks as follows:

a. $\psi_5$
b. $\frac{1}{\sqrt{2}}(\psi_2 + \psi_3)$
c. $\frac{1}{\sqrt{2}}(\psi_1 - \psi_4)$
d. $\sqrt{\frac{1}{5}}\psi_1 + \sqrt{\frac{4}{5}}\psi_3$

To answer this question correctly, students need to recognize that the frequency of the oscillation is proportional to the energy difference between the superposition terms. In our sample population, all students had seen mathematical calculation of such an oscillation frequency for time evolution of superposition states; however, very few were able to make the connection between the energy difference and oscillation frequency.

### Example 2: Operators & Hamiltonian

As another example, we consider open-ended QMAT question twelve (Q12). The wording of the question was not altered in the MC version (Q18 in **Fig. 3**), which asks whether a system in an eigenstate of an arbitrary operator will remain in that eigenstate until disturbed by measurement. The average score for open-ended QMAT Q12 was very low (CPP=23%, N=19; CUB=25%, N=53). Using student written responses we constructed the distractors in MC-Q19.

While many students correctly identified the false nature of the statement in MC-Q18, (CPP = 71%, CUB = 52%), many were not able to detect all of the possible correct options (d, e, f) in Q19. (CPP = 18%, CUB = 15%). Item analysis showed that option (f) for Q19 was very unpopular (only one student in both data sets selected option (f)), despite being a correct answer choice. In student interviews there were frequent comments about the novelty of a time-dependent Hamiltonian and the lack of intuition about the behavior of such a system, thus option (f) was removed in later versions of the test. Furthermore, MC questions with multiple correct answer choices showed undesirable item difficulty (0% of students in both samples selected all correct options and no incorrect options), so options (d) and (e) were combined in later versions of the instrument.

---

**Q18.** Is the following statement true or false for all operators $\hat{Q}$? A system which is in an eigenstate of $\hat{Q}$ will stay in that state until disturbed by measurement.
  a. True
  b. False

**Q19.** What is the reasoning for your answer to the previous question? Choose all that apply.
  a. The system remains in an eigenstate of operator $\hat{Q}$ until measurement of a non-commuting operator.
  b. The expectation value of $\hat{Q}$ is time-independent.
  c. The system will evolve and the exact state cannot be known.
  d. The Hamiltonian may not commute with $\hat{Q}$.
  e. The operator $\hat{Q}$ may have a time dependence.
  f. The Hamiltonian may have a time dependence.

---

**Figure 3**. The first version of two MC questions corresponding to open-ended QMAT Q12.

## Expert Validation

Validity is a measure of whether or not the test measures what it says it measures. The face and content validity of the open-ended test items was established in the original development process of the QMAT [10]. Expressing common student incorrect ideas in a clear and concise language that would make sense to experts is a challenging task in general. Especially for abstract topics of QM, absence of intuition or real world experience limits forming many preconceptions that can be easily classified. In working to establish clear language to expose student common ideas, the project team (two faculty members who have taught the course and two undergraduates who have completed the courses with outstanding performances) held a series of meetings to probe all questions and distractors for clarity and content. The questions and the distractors were reworded and iteratively enhanced.

For further content-related validity for the MC version, we examined how well the test items cover the content domain it purports to test and how well the distractors represent specific student ideas. A CPP faculty member, who has taught the course several times in the past and has written several textbooks, including one on Quantum Theory, reviewed the MC questions. After careful examination of each item and along with constructive feedback, he indicated that the test *"... is [a] comprehensive set that is capable of probing a variety of conceptual hurdles and difficulties commonly encountered by beginning students of quantum mechanics."* Additional feedback was obtained from the original QMAT author, and through interviews and feedback from the current course instructors at CPP and CUB.

## DISCUSSION

Assessment is central to teaching and learning as it provides information about the gap between what students currently know and what they need to know, and serves as a tool to systematically address improving pedagogy. Nevertheless, test development, even from an existing instrument, is a long and difficult process that often requires several iterations.

We discussed here examples of mechanisms by which QMAT questions were converted to MC and iterated upon through the use of data. We discussed some challenges and affordances of converting the test to MC. Much of the success of MC tests is attributed to the careful construction of each question, as well as each response [22-23]. Thus, we made use of long answer responses to inform the development of MC items and distractors. Open-ended questions have some advantages over MC format when it comes to making student thinking more visible; nonetheless, a high quality MC instrument can still enable us to learn about student ideas. While this new multiple-choice instrument is still under refinement and further development, we hope that it will address important yet challenging areas in upper division courses. We will be testing the modified version and will conduct further statistical analyses of reliability and validity. To make this test t equally useful for traditional and modern approaches in teaching QM, in future versions, we are considering explicitly including questions in the context of discrete spin one-half.

## ACKNOWLEDGMENTS

We acknowledge valuable feedback of faculty, especially Kai Lam, and Cindy Regal. We gratefully acknowledge the work of Steven Goldhaber, who originally developed the QMAT and provided invaluable feedback. We are also grateful for student willingness to participate in this study and interviews.